\documentclass[preprint,12pt]{elsarticle}

\usepackage{amssymb}

\usepackage{amsmath}
\usepackage{subfigure}
\usepackage{bm}
\usepackage{url}
\usepackage{color}
\usepackage{here}

\usepackage{lineno}
\usepackage{setspace}
\journal{Building and Environment}

\begin{document}

\begin{frontmatter}



\title{Super-Resolution of Three-Dimensional Temperature and Velocity for Building-Resolving Urban Micrometeorology Using Physics-Guided Convolutional Neural Networks with Image Inpainting Techniques}


\affiliation[inst1]{
    organization={Global Scientific Information and Computing Center, Tokyo Institute of Technology},
    addressline={2-12-1 Ookayama, Meguro-ku},
    city={Tokyo},
    postcode={1528550},
    country={Japan}
}

\affiliation[inst2]{
    organization={Research Institute for Value-Added-Information Generation, Japan Agency for Marine-Earth Science and Technology},
    addressline={3173-25 Showa-machi, Kanazawa-ku, Yokohama},
    city={Kanagawa},
    postcode={2360001},
    country={Japan}
}

\author[inst1]{Yuki Yasuda\fnref{corresponding_authoer_info}}
\ead{yasuda.y.aa@m.titech.ac.jp}
\fntext[Corresponding_authoer_info]{Corresponding author. 2-12-1-i9-111 Ookayama, Meguro-ku, Tokyo, 1528550, Japan. Tel.: +81 3 57343173 (office). Email address: yasuda.y.aa@m.titech.ac.jp}

\author[inst1]{Ryo Onishi}
\ead{onishi.ryo@gsic.titech.ac.jp}

\author[inst2]{Keigo Matsuda}
\ead{k.matsuda@jamstec.go.jp}

\begin{abstract}
Atmospheric simulations for urban cities can be computationally intensive because of the need for high spatial resolution, such as a few meters, to accurately represent buildings and streets. Deep learning has recently gained attention across various physical sciences for its potential to reduce computational cost. Super-resolution is one such technique that enhances the resolution of data. This paper proposes a convolutional neural network (CNN) that super-resolves instantaneous snapshots of three-dimensional air temperature and wind velocity fields for urban micrometeorology. This super-resolution process requires not only an increase in spatial resolution but also the restoration of missing data caused by the difference in the building shapes that depend on the resolution. The proposed CNN incorporates gated convolution, which is an image inpainting technique that infers missing pixels. The CNN performance has been verified via supervised learning utilizing building-resolving micrometeorological simulations around Tokyo Station in Japan. The CNN successfully reconstructed the temperature and velocity fields around the high-resolution buildings, despite the missing data at lower altitudes due to the coarseness of the low-resolution buildings. This result implies that near-surface flows can be inferred from flows above buildings. This hypothesis was assessed via numerical experiments where all input values below a certain height were made missing. This research suggests the possibility that building-resolving micrometeorological simulations become more practical for urban cities with the aid of neural networks that enhance computational efficiency.
\end{abstract}




\end{frontmatter}


\section{Introduction} \label{sec:introduction}

Global urbanization indicates that various challenges in cities, such as reducing heat stress and energy consumption, will become more important in the near future \citep{WCR2020}. The estimation of environmental conditions such as wind and temperature may be useful for addressing these tasks \citep{Mauree+2019RSER, Lai+2019STTE}. Indeed, architects and urban planners are now incorporating the knowledge of urban micrometeorology, e.g., thermal and dynamical airflow responses, into building designs \citep{Naboni2013PLEA, Toparlar+2017RSER, Lam+2021SCS}. Numerical simulation based on the laws of physics is an effective tool for estimating atmospheric states on micrometeorological scales \citep{Toparlar+2017RSER, Lam+2021SCS}. However, these simulations can be computationally intensive due to the need for high spatial resolution such as a few meters to represent buildings and streets. Therefore, the reduction in computational cost is necessary to make micrometeorological simulations more practical for various applications in urban areas.

Recently, super-resolution (SR) has been proposed as a promising approach for reducing computational cost in numerical simulations across various physical sciences \citep[e.g.,][]{Onishi+2019SOLA, Ramanah+2020MNRAS, Shiina+2021SR}. Super-resolution is a technique that enhances the resolution of images and has been focused on in computer vision as an application of neural networks (NNs) \citep{Dong2014ECCV, Ha2019, Anwar+2020ACMCS}. The success of such NNs has led to numerous applications of SR for various fluid systems \citep{Ducournau+2016PRRS, Xie+2018ACM, Deng+2019PF, Fukami+2019JFM, Fukami+2021JFM, Wang+2021PF, Sales+2021PACT, Sales+2022arXiv, Jiang+2020IEEE, Wang+2020NIPS, Bode+2021PCI, Bao+2022CUAI}, including atmospheric flows \citep{Onishi+2019SOLA, Wu+2021GRL, Wang+2021GMD, Vandal+2017ACM, Rodrigues+2018IEEE, Stengel+2020PNAS, Yasuda2022BAE}. The recent progress in the fluid SR is summarized in a review paper \citep{Fukami+2023arXiv}. One application in meteorology is SR simulation \citep{Onishi+2019SOLA, Wu+2021GRL, Wang+2021GMD}, where the time evolution is calculated from low-resolution (LR) atmospheric simulations, and high-resolution (HR) inferences are obtained by inputting the LR results into a trained NN. SR simulation is computationally efficient because it only requires LR atmospheric simulations and the elapsed time of NN inference is usually negligible.

The potential of SR simulation has been investigated for urban cities \citep{Onishi+2019SOLA, Wu+2021GRL, Yasuda2022BAE}. These studies focused on the SR of two-dimensional fields such as temperature. They reported that the multiple input improves the validity of inference because NNs can learn the physical relationships between these variables \citep{Wu+2021GRL, Yasuda2022BAE}. For more practical applications, however, it is necessary to super-resolve three-dimensional fields such as wind velocity.

It may be challenging to super-resolve micrometeorological simulation results in cities because the streets between buildings may not be resolved at LR. In such simulations, the atmospheric flows outside buildings are represented, while the grid points inside buildings are missing. Thus, the SR process requires not only an increase in spatial resolution but also the reconstruction of flows around buildings to account for changes in building shape from LR to HR. Our previous studies on the two-dimensional SR \citep{Onishi+2019SOLA, Yasuda2022BAE} avoided this issue by sampling temperature at a certain height above the bottom surface such as the ground or roofs of buildings. As a result of this sampling, the two-dimensional data contained no missing value. Furthermore, in other fluid systems, such as flows around airfoils \citep{Wang+2021PF, Sales+2021PACT, Sales+2022arXiv} and cylinders \citep{Deng+2019PF, Fukami+2019JFM, Fukami+2021JFM, Wang+2021PF}, the changes in obstacle shape during the SR process have not been considered. Therefore, it is currently unclear how to super-resolve flows around buildings whose shape depends on the resolution.

In computer vision, the reconstruction of missing pixels is referred to as image inpainting, and deep-learning based methods have been studied \citep{Elharrouss+2020NPL, Qin+2021Display}. For instance, missing pixels are considered masked and are recovered based on the surrounding pixels or statistical information learned from data. These techniques have been applied to various observation data, including satellite imagery, to fill in missing values caused by various factors such as cloud cover, measurement error, or data sparsity \citep{Li+2019JPRS, Kadow+2020NG, Meraner+2020JPRS, Geiss+2021AMT, Maduskar+2021AISC, Schweri+2021FC, Czerkawski+2022RS}. However, these methods cannot be directly applied to urban micrometeorological data because missing regions remain after the conversion from LR to HR. In other words, HR atmospheric flows are still considered masked by HR buildings.

The changes in the shape of missing regions may be addressed by introducing two masks at LR and HR. However, there are few studies utilizing multiple masks. Schweri et al. \citep{Schweri+2021FC} developed a U-Net \citep{Ronneberger+2015MICCAI} by using two types of masks to reconstruct the atmospheric vortex street, which is missed due to cloud cover, behind Guadalupe Island. The first mask, called the obstacle mask, represents obstacles (i.e., Guadalupe Island) that remain in the output. The other mask, called the inpainting mask, corresponds to cloud cover and is varied in partial convolution \citep{Liu+2018ECCV}. The inpainting mask takes 0 at missing pixels and 1 at non-missing pixels. In the partial convolution \citep{Liu+2018ECCV}, missing pixels are restored by convolution based only on the surrounding non-missing pixels, while the inpainting mask is shrunk according to this restoration. Although the U-Net developed by Schweri et al. \citep{Schweri+2021FC} successfully restored missing data of horizontal flows around obstacles, they did not perform SR or consider three-dimensional flows. To the authors' knowledge, the combination of SR and image inpainting has yet to be investigated for turbulent flows in three dimensions.

This paper proposes a convolutional neural network (CNN) that super-resolves three-dimensional air temperature and wind velocity fields in urban cities. The SR process in the CNN accounts for the building-shape difference between LR and HR. This CNN employs an image inpainting technique called gated convolution \citep{Yu+2019ICCV}, which is an extension of the partial convolution \citep{Liu+2018ECCV}. The CNN performance was evaluated by utilizing building-resolving micrometeorological simulations around Tokyo Station in Japan. Specifically, the LR input data were created from the HR simulation results and used to train the CNN via supervised learning. The training loss function included a physics loss term, namely the mean divergence loss, to improve the validity of inference. This proof of concept demonstrates the feasibility of enhancing spatial resolution while simultaneously reconstructing flows around obstacles. This is an essential step toward the development of SR simulation for urban micrometeorological predictions.

The remaining paper is organized as follows. Section \ref{sec:cnn} presents the CNN for the SR of atmospheric flows around buildings. Section \ref{sec:methods} describes the methods to train and evaluate the CNN. The results are analyzed in Section \ref{sec:results-discussion}. We discuss the possibility of qualitatively inferring near-surface flows from upper flows. Finally, the conclusions are given in Section \ref{sec:conclusions}.

\section{CNN for super-resolution accounting for building-shape changes} \label{sec:cnn}

The proposed CNN enhances the spatial resolution of three-dimensional air temperature and wind velocity fields on Cartesian coordinates, while simultaneously modifying the missing regions associated with changes in building shape from LR to HR. The input consists of five quantities: LR temperature $T^{\rm LR}$ and velocity $(u^{\rm LR}, v^{\rm LR}, w^{\rm LR})$, together with HR building mask $B^{\rm HR}$. The output is composed of four quantities: HR temperature $T^{\rm HR}$ and velocity $(u^{\rm HR}, v^{\rm HR}, w^{\rm HR})$. Here, the eastward, northward, and upward components of the velocity vector $\bm{V}$ are represented by $u$, $v$, and $w$, respectively. In the following, the superscripts of $\rm LR$ and $\rm HR$ are omitted if there is no confusion. The temperature and velocity are missing at the grid points inside buildings. All input and output are instantaneous snapshots and are described as three-dimensional numerical arrays that correspond to the geometrical distributions in the three-dimensional space.

The building mask $B$ is a binary field that takes 1 and 0 at the grid points outside and inside buildings, respectively:
\begin{equation}
    B = \begin{cases}
        1 & (\text{outside buildings}), \\
        0 & (\text{inside buildings}).
    \end{cases}
\end{equation}
We assume that buildings are described by voxels and each grid point is located in the center of the voxel, i.e., it is not on building walls or roofs. Further, the building mask $B$ is assumed to be static. By element-wise multiplication, $B$ can extract non-missing values outside buildings from any field quantity. Only the HR building mask is utilized, and the LR mask is not required, as explained below.

\begin{figure}[H]
    \centering
    \includegraphics[width=16cm]{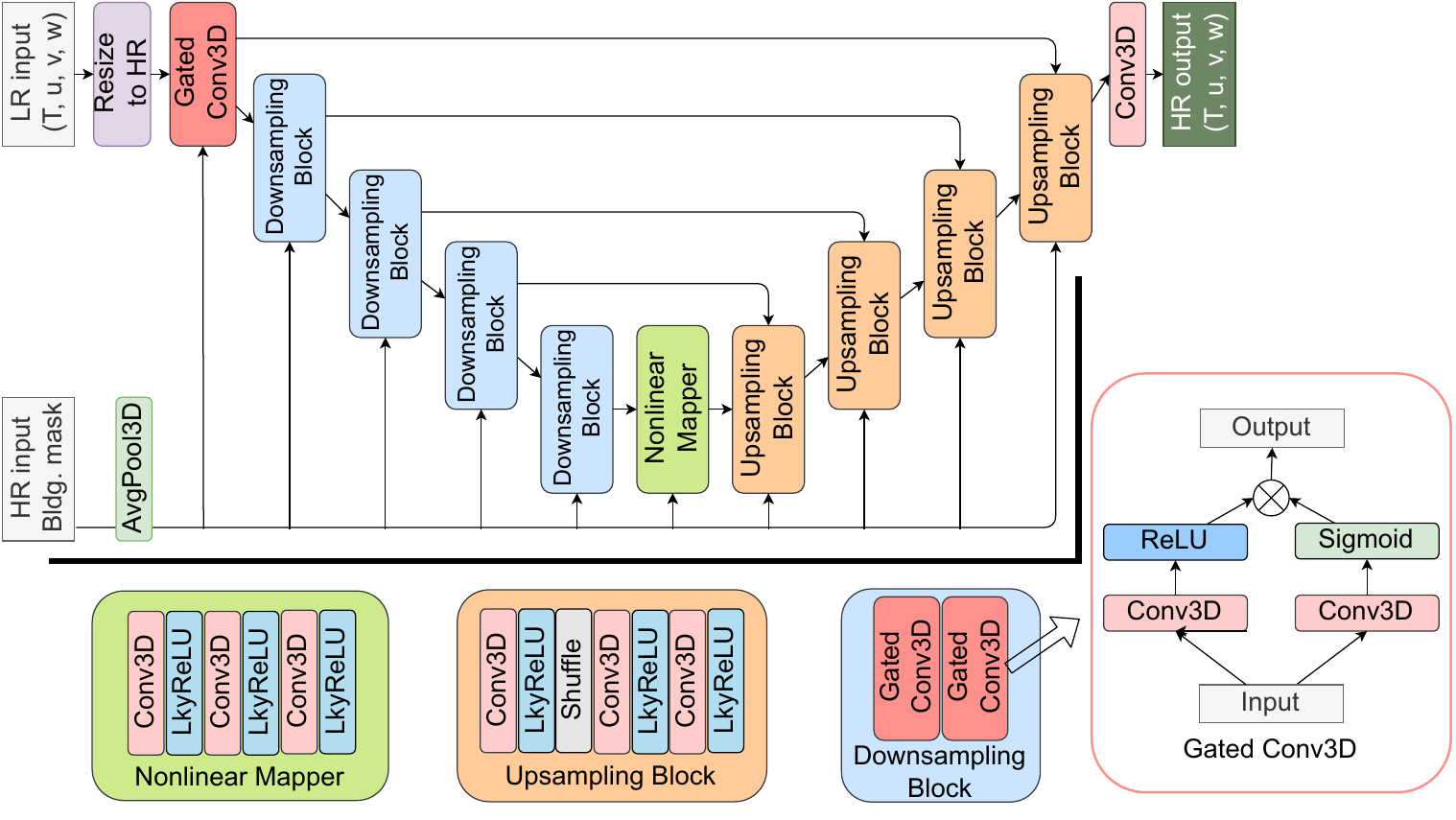}
    \caption{Architecture of the convolutional neural network (CNN) proposed in this study. Details of Nonlinear Mapper, Upsampling Block, Downsampling Block, and Gated Conv3D are displayed in the lower panels. The label ``AvgPool3D'' is three-dimensional average pooling, ``Conv3D'' is three-dimensional convolution, ``LkyReLU'' is Leaky ReLU \citep{Maas+2013ProcICML}, ``Shuffle'' means the three-dimensional version of pixel shuffle \citep{Shi+2016CVPR}, and $\otimes$ represents element-wise multiplication. If two or three arrows point to a block, all inputs are concatenated and then fed to the first layer in the block. The detailed implementation is available on the Zenodo repository (see Data availability).}
    \label{fig:cnn-architecture}
\end{figure}

Figure \ref{fig:cnn-architecture} shows the network architecture of the proposed CNN, which is based on the U-Net developed by Schweri et al. for image inpainting \citep{Schweri+2021FC}. Hyperparameters such as kernel size were tuned through a grid search. The detailed implementation, including the hyperparameters, can be found on the Zenodo repository (see Data availability). The calculation process in the CNN consists of four steps. First, the LR input of $T^{\rm LR}$, $u^{\rm LR}$, $v^{\rm LR}$, and $w^{\rm LR}$ is aligned in size with the HR output by using the nearest neighbor interpolation. Second, the resized input is downsampled with gated convolution \citep{Yu+2019ICCV}, which is an image inpainting technique described in the next paragraph. In each downsampling block, the first gated convolution has a stride of $2 \times 2 \times 2$, halving the array size in each dimension. Third, the downsampled input is transformed by ordinary convolution and Leaky ReLU \citep{Maas+2013ProcICML}. Fourth, the upsampling blocks restore the array size to the HR. In each upsampling block, the output from the corresponding downsampling block is passed through a skip connection \citep{Ronneberger+2015MICCAI, He+2016CVPR}. Upsampling is then performed by using the three-dimensional version of pixel shuffle \citep{Shi+2016CVPR}, where the array size is doubled in each dimension. Importantly, the HR building mask is resized with average pooling where necessary and fed to each downsampling or upsampling block \citep{Schweri+2021FC}.

In the gated convolution \citep{Yu+2019ICCV}, the features are multiplied with the weights calculated by a sigmoid function (Fig. \ref{fig:cnn-architecture}). This operation can be viewed as an extension of the partial convolution \citep{Liu+2018ECCV}. The partial convolution is rule-based, where only non-missing values are convolved using a prescribed mask. In contrast, the gated convolution is expected to optimize the mask, namely the weights for non-missing values, depending on the input via the sigmoid function.

The following preprocessing is applied to regard zeros as missing values. Each variable $X$ ($= T^{\rm LR}$, $u^{\rm LR}$, $v^{\rm LR}$, or $w^{\rm LR}$) is transformed as
\begin{equation}
     {\rm clip}_{[0, 1]}\left( \frac{X-m_X}{s_X}\right). \label{eq:preprocess}
\end{equation}
The clipping function, ${\rm clip}_{[0,1]}(x) = \min\{1, \max\{0, x\}\}$ ($x \in \mathbb{R}$), limits the value range to $[0, 1]$. The parameters $m_X$ and $s_X$ are determined such that 99.9\% of $X$ values are within the range of $[0,1]$. After applying (\ref{eq:preprocess}), NaN (Not a Number) values at the grid points inside buildings are replaced with zeros. Due to this preprocessing, the spatial distribution of zeros in the input represents the LR buildings, and the LR building mask is not required.

\section{Methods} \label{sec:methods}

\subsection{Building-resolving micrometeorological simulations} \label{subsec:micrometeorology}

Building-resolving micrometeorological simulations were conducted by employing the Multi-Scale Simulator for the Geoenvironment (MSSG) \citep{Takahashi+2013JPCS, Onishi+2012JAS, Sasaki+2016GRL, Matsuda+2018JWEIA}. MSSG is a coupled atmosphere-ocean model that operates on global, meso-, and urban scales. The atmospheric component of MSSG, i.e., MSSG-A, can be utilized as a building-resolving large eddy simulation (LES) model in conjunction with a three-dimensional radiative transfer model \citep{Matsuda+2018JWEIA}. The governing equations of MSSG-A include the conservation equations of mass, momentum, and energy for compressible flows and the transport equations for mixing ratios of water substances including water vapor, liquid, and ice cloud particles. Further information on the configuration can be found in our precursor studies \citep{Onishi+2019SOLA, Matsuda+2018JWEIA}.

\begin{figure}[H]
    \centering
    \includegraphics[width=16cm]{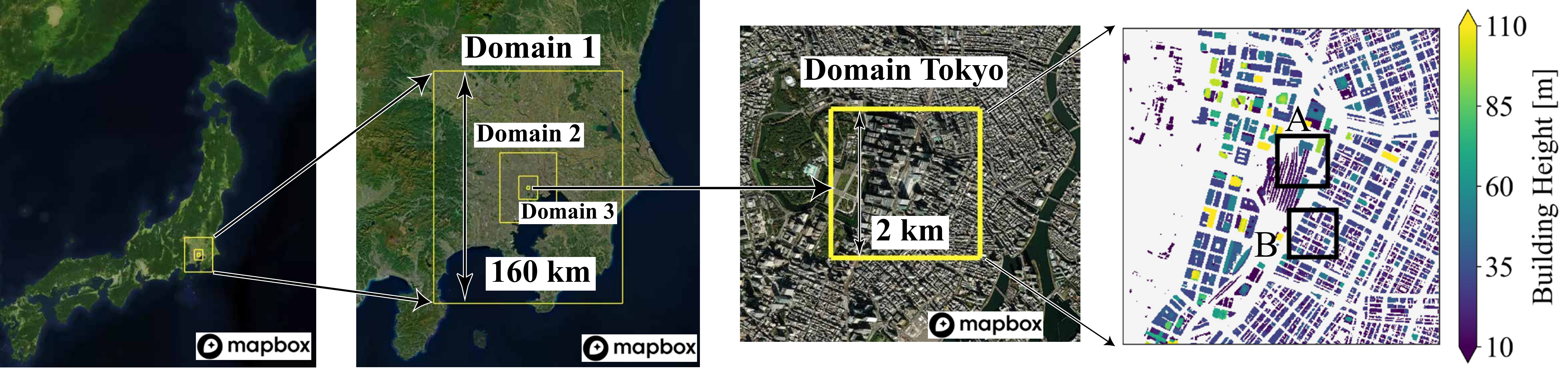}
    \caption{Computational domains for the mesoscale and micrometeorological simulations. The latter resolves the buildings around Tokyo Station, as shown in the rightmost building height distribution. The satellite images were obtained from MapBox \citep{mapbox} and OpenStreetMap \citep{openstreetmap}. In the building height distribution, the upper black square, labeled ``A'', represents the region shown in Fig. \ref{fig:2m-height-temperature}, and the lower square, labeled ``B'', represents the region shown in Figs. \ref{fig:bldg-height}, \ref{fig:sr-057m-z0-b}, and \ref{fig:sr-057m-z2-b}.}
    \label{fig:computation-domains}
\end{figure}

Dynamical downscaling \citep{Giorgi+2015} was performed for an actual urban area around Tokyo Station in Japan (35.6809$^\circ$N and 139.7670$^\circ$E) to reduce the spatio-temporal scales of atmospheric flows from meso- to urban scales. The mesoscale simulations used three two-way-coupled nested systems (Fig. \ref{fig:computation-domains}): Domain 1 through 3. These domains are characterized by horizontal grid points of $160 \times 160$, with grid spacings of $1$ km for Domain 1, $300$ m for Domain 2, and $100$ m for Domain 3. The vertical grids for Domains 1 through 3 are identical and consist of 55 points over an altitude range of $40$ km. The computation domain of micrometeorology, i.e., Domain Tokyo (Fig. \ref{fig:computation-domains}), is nested within Domain 3. This smallest domain has horizontal grid points of $400 \times 400$ with a resolution of 5 m, as well as 151 stretched grid points in the vertical dimension. The vertical grid spacings are 5 m below an altitude of 500 m and varied between 500 and 1500 m. The building height distribution was created from the geographic information system (GIS) data for 2011 provided by the Tokyo Metropolitan Government. The created distribution has a resolution of 5 m, which is sufficiently high to represent streets between buildings (Fig. \ref{fig:computation-domains}). The initial and boundary conditions for the mesoscale simulations were obtained from grid-point-value data \citep{jmbsc} of the Meso-Scale Model at the Japan Meteorological Agency, whereas those for the micrometeorological simulations were obtained from the results of Domain 3.

\begin{figure}[H]
    \centering
    \includegraphics[width=9cm]{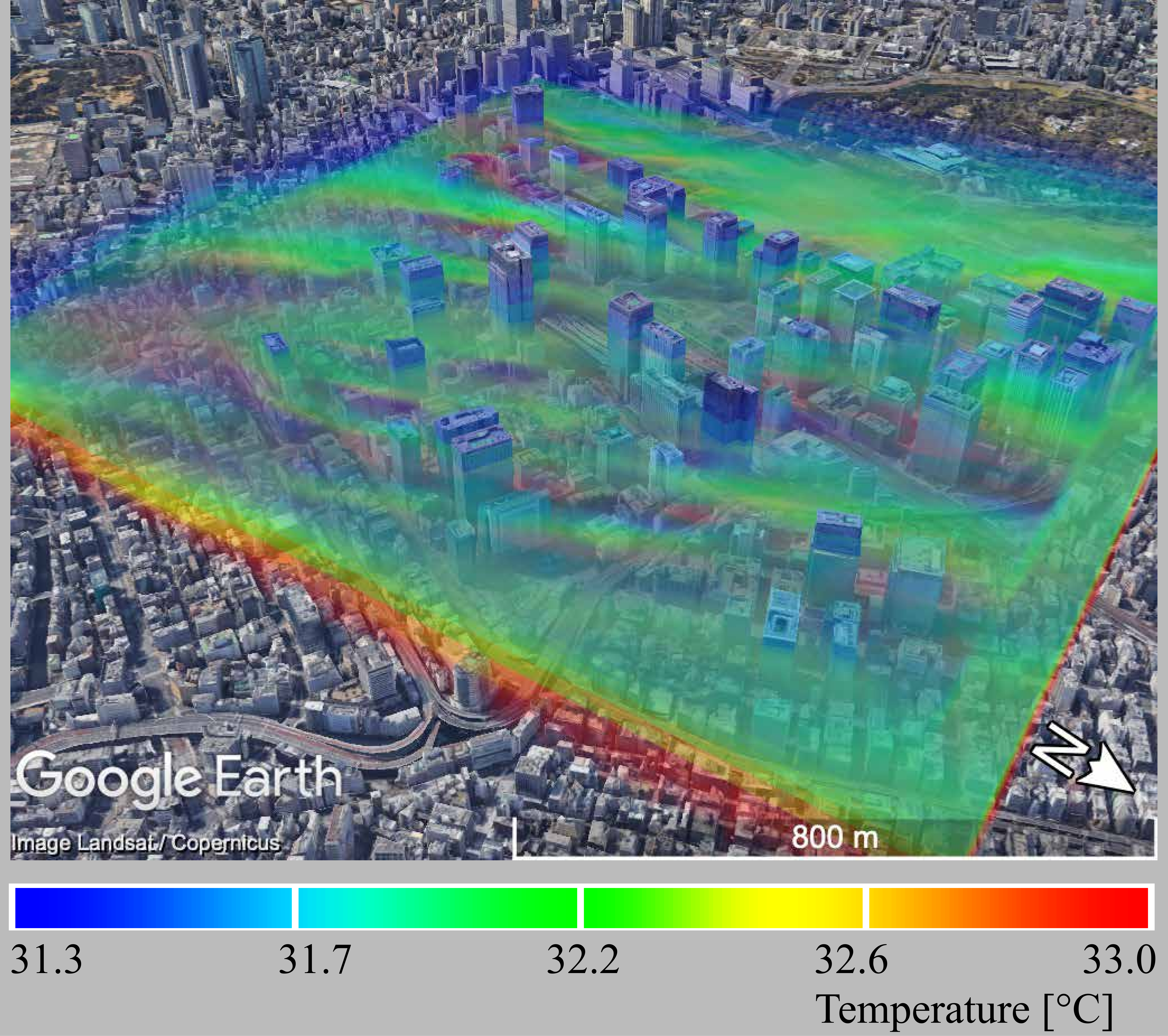}
    \caption{A three-dimensional distribution of the 1-minute average temperature at 2013-07-09 14:00+09:00 from the micrometeorological simulation. The Volume Data Visualizer for Google Earth (VDVGE) \citep{Kawahara+2015SIGGRAPH} was utilized for the visualization.}
    \label{fig:vdvge}
\end{figure}

The numerical experiments were conducted for hot days between 2013 and 2020. The focus on hot days was motivated by the possibility that the risk of heat-related illness could be influenced by temperature on street scales \citep{Kamiya+2019IJERPH}. Specifically, 171 hot summer hours were selected between 2013 and 2020 in which the maximum daily temperature exceeded 35$^\circ$C. Figure \ref{fig:vdvge} shows an instantaneous snapshot of the simulated three-dimensional temperature, which appears to be advected by the flows passing around the buildings. Each simulation was run for each targeted hour. The result from the first 10 minutes was discarded, and the remaining 50 minutes were used to obtain 1-minute average values. These average fields were sampled every 2 minutes. This sampling interval was determined by the decrease in auto-correlation below a threshold of 0.5. Consequently, 25 snapshots in three dimensions were obtained from each experiment, totaling 4,275 snapshots ($= 25 \times 171$).

\subsection{Data preparation for deep learning}

Pairs of HR and LR data were generated for supervised learning. First, we cropped the HR results from the micrometeorological simulations as follows. The central $320 \times 320$ grid points were extracted to eliminate the influence of the lateral damping layer for dynamical downscaling; then, vertically, the bottom 32 grid points were selected to focus on atmospheric flows near the ground. As a result, each HR snapshot consists of $320 \times 320 \times 32$ voxels. Since the resolution is 5 m in all directions, the east-west and north-south widths are 1,600 m and the vertical width is 160 m.

\begin{figure}[H]
    \centering
    \includegraphics[width=9cm]{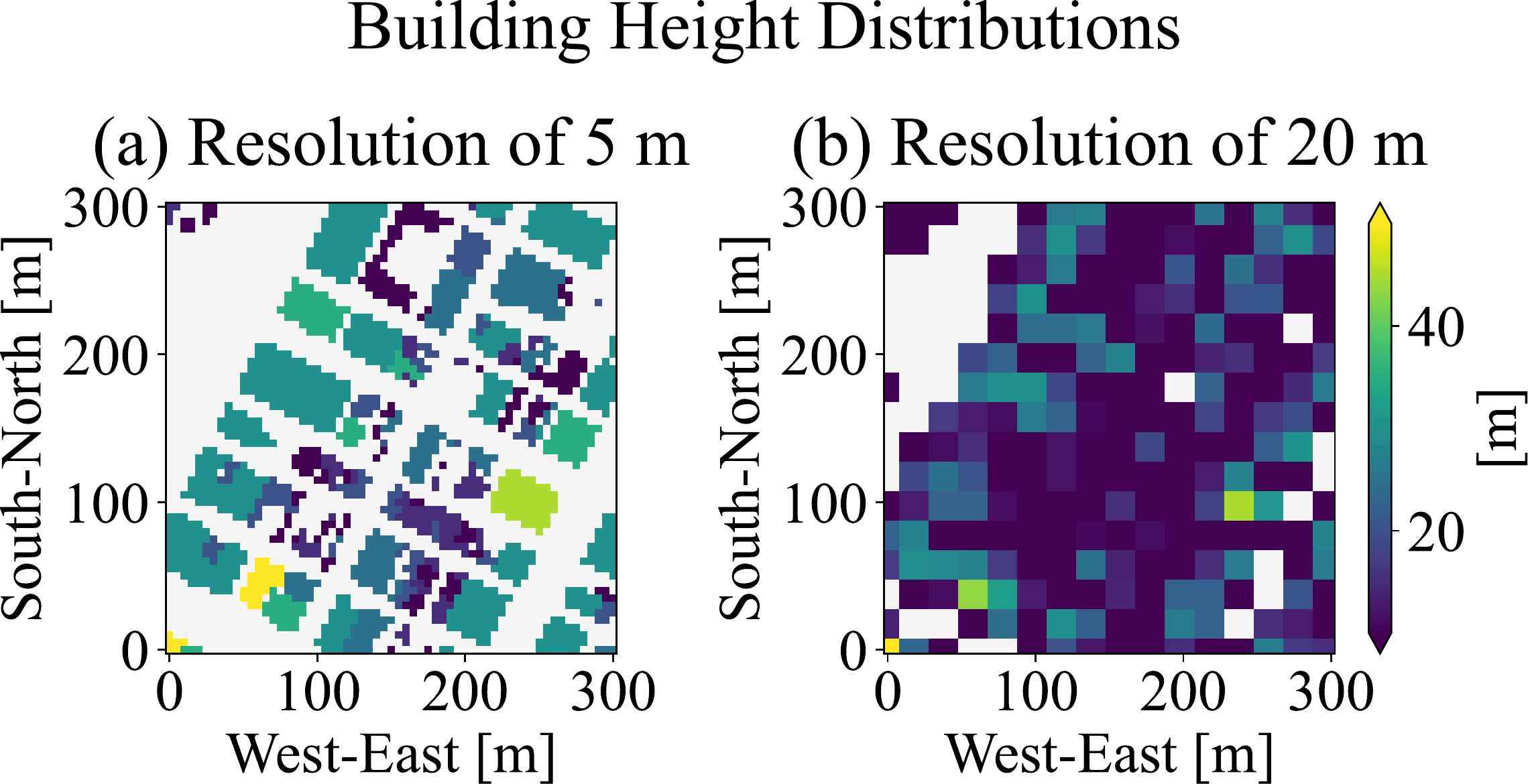}
    \caption{Building height distributions over a cropped region: (a) 5 m and (b) 20 m resolution. This region is labeled ``B'' in Fig. \ref{fig:computation-domains}.}
    \label{fig:bldg-height}
\end{figure}

The LR data were obtained by applying the average pooling to the HR data such that the resolution is four times lower, i.e., the 20 m resolution in all directions. The factor of four is sufficiently coarse for practical applications \cite{Onishi+2019SOLA, Wang+2021GMD, Yasuda2022BAE}. The LR building height distribution was obtained from the HR distribution. Specifically, the entire $320 \times 320$ region was divided into $4 \times 4$ blocks, and each block was replaced with the mean value. This operation is called the average pooling with a $4 \times 4$ kernel. The total volume of buildings is preserved between the HR and LR, even though some streets between buildings are not represented at the LR, as shown in Fig. \ref{fig:bldg-height}. This fact follows that the total volume of the atmosphere is preserved as well. The LR temperature and velocity were then created in two steps. First, the average pooling with a $4 \times 4 \times 4$ kernel was applied to each HR snapshot ($T^{\rm HR}$, $u^{\rm HR}$, $v^{\rm HR}$, or $w^{\rm HR}$). In this average operation, the grid points within the LR buildings were excluded. Second, NaN values were inserted at the grid points of the LR buildings. In the subsequent preprocessing, these NaN values were replaced with zeros (Section \ref{sec:cnn}). The resultant LR snapshots consist of $80 \times 80 \times 8$ voxels and are four times coarser than the HR snapshots.

All pairs of HR and LR data were split into training, validation, and test datasets in proportions of 60\%, 20\%, and 20\%, respectively. This splitting preserved the chronological order to prevent the so-called data leakage. In more detail, the training dataset was obtained from 99 micrometeorological simulations between 2013 and 2018 (2,475 pairs of HR and LR snapshots); the validation set was from 36 simulations between 2018 and 2019 (900 pairs); and the test set was from 36 simulations between 2019 and 2020 (900 pairs).

\subsection{Training of the CNN}

The CNN was trained via supervised learning using the following loss function:
\begin{align}
    \| \bm{{Y}^{\rm HR}} - \bm{\hat{Y}^{\rm HR}} \|^2 &+ \lambda_{\rm grd} \| {B}^{\rm HR} \odot \nabla(\bm{Y^{\rm HR}} - \bm{\hat{Y}^{\rm HR}}) \|^2 \notag \\ &+ \lambda_{\rm div} \| {B}^{\rm HR} \odot (\nabla\cdot\bm{V^{\rm HR}} - \nabla\cdot\bm{\hat{V}^{\rm HR}}) \|^2, \label{eq:loss}
\end{align}
where $\bm{Y^{\rm HR}}$ represents the ground truth, $\bm{Y^{\rm HR}} = (T^{\rm HR}, u^{\rm HR}, v^{\rm HR}, w^{\rm HR}) = (T^{\rm HR}, \bm{{V}^{\rm HR}})$; $\bm{\hat{Y}^{\rm HR}}$ is the corresponding inference by the CNN; $\lambda_{\rm grd}$ and $\lambda_{\rm div}$ are positive real constants; $\lVert \cdot \rVert^2$ stands for mean squared quantities; $\odot$ is element-wise multiplication; and $\nabla$ denotes the second-order centered difference. The first term in (\ref{eq:loss}) is the mean squared error. The second term is the mean gradient error \citep{Schweri+2021FC, Lu+2022SIVP}. The third term is the mean divergence error \citep{Bode+2021PCI, Bao+2022CUAI, Schweri+2021FC}. The building mask ${B}^{\rm HR}$ is included in the second and third terms to exclude the invalidity of finite differences at the grid points near building boundaries. The first term assesses the flow-field values, whereas the second and third terms measure the difference in the flow-field smoothness. Previous studies reported the effectiveness of the second or third term \citep{Bode+2021PCI, Bao+2022CUAI, Schweri+2021FC, Lu+2022SIVP}. In particular, the third term, i.e., the error in the divergence of three-dimensional velocity, can enhance the physical validity of inference. Due to this term, the trained CNN is regarded as a physics-guided NN.

The CNN was trained by using the Adam optimizer \citep{Kingma+2015ICLR} with a learning rate of $1.0 \times 10^{-4}$. Here, the parameters in the loss function (\ref{eq:loss}) were set at $\lambda_{\rm grd} = 1$ and $\lambda_{\rm div} = 10$. We stored the weights of the CNN that resulted in the smallest validation loss after 1,200 epochs. During training, $64 \times 64 \times 32$ voxels were randomly cropped from the HR data, and the corresponding region was extracted from the LR data. Repeating this sampling, a mini-batch of 32 samples was created and fed to the CNN.

The CNN was implemented with PyTorch 1.11.0 \citep{pytorch2019} and trained with distributed data parallel (DDP) using two GPU boards of NVIDIA A100 40GB PCIe on the Earth Simulator supercomputer system in Japan Agency for Marine-Earth Science and Technology. During the evaluation, one GPU was used. The CNN implementation is available on the Zenodo repository (see Data availability).

\subsection{Evaluation metrics for the CNN}

The CNN performance was evaluated using two types of metrics for voxel-wise accuracy and pattern consistency, where temperature and velocity were analyzed separately. At the evaluation phase, the inference was obtained by feeding LR snapshots of the full size $80 \times 80 \times 8$ into the trained CNN.

The mean voxel-wise errors in temperature and velocity are defined as follows:
\begin{align}
    \left\lvert\Delta T\right\rvert &= \frac{1}{N} \sum_{\text{const height}}  B^{\rm HR}\left\lvert T^{\rm HR} - \hat{T}^{\rm HR} \right\rvert, \notag \\
    &= \frac{1}{N} \sum_{\text{const height}}  B^{\rm HR}\sqrt{\left(T^{\rm HR} - \hat{T}^{\rm HR}\right)^2}, \label{eq:mae-T}
\end{align}
and
\begin{align}    
    \left\lvert\Delta \bm{V}\right\rvert &= \frac{1}{N} \sum_{\text{const height}}  B^{\rm HR} \left\lvert \bm{V^{\rm HR}} - \bm{\hat{V}^{\rm HR}} \right\rvert, \notag \\
    &= \frac{1}{N} \sum_{\text{const height}}  B^{\rm HR} \sqrt{\left(u^{\rm HR} - \hat{u}^{\rm HR}\right)^2 + \left(v^{\rm HR} - \hat{v}^{\rm HR}\right)^2 + \left(w^{\rm HR} - \hat{w}^{\rm HR}\right)^2 }, \label{eq:mae-V}
\end{align}
where the summation is taken at a constant height over the horizontal sections in all test data, and $N$ denotes the number of the grid points outside buildings, i.e., $N = \sum_{\text{const height}} B^{\rm HR}$. Equations (\ref{eq:mae-T}) and (\ref{eq:mae-V}) are referred to as the error norm of temperature and velocity, respectively.

The pattern consistency is estimated with the mean structural similarity index measure (MSSIM) \citep{Wang+2004IEEE}:
\begin{align}
    \text{MSSIM loss} &= 1 - {\rm MSSIM}, \notag \\ &= 1 - \frac{1}{N}\sum_{\text{const height}} \left[\frac{\left(2\mu\hat{\mu} + {\rm C_1} \right) \left(2\chi + {\rm C_2} \right)}{\left(\mu^2 + \hat{\mu}^2 + {\rm C_1}\right) \left( \sigma^2 + \hat{\sigma}^2 + {\rm C_2} \right)} \right], \label{eq:mssim}
\end{align}
where ${\rm C_1} = 0.01$, ${\rm C_2} = 0.03$, $\mu$ and $\sigma^2$ are, respectively, the mean and variance of the ground truth; $\hat{\mu}$ and $\hat{\sigma}^2$ are the corresponding quantities of the inference; and $\chi$ is the covariance between the ground truth and inference. The variables in the summation are calculated locally by applying a three-dimensional Gaussian filter along with the building mask $B^{\rm HR}$. The MSSIM loss takes a value larger than or equal to $0$, and smaller values indicate that the spatial patterns of inference are more similar to those of the ground truth. A detailed discussion on MSSIM is found in Wang et al. \citep{Wang+2004IEEE}. Equation (\ref{eq:mssim}) was applied directly to temperature fields. For velocity, (\ref{eq:mssim}) was applied to each component of $u$, $v$, and $w$, and the obtained three MSSIM losses were averaged. We confirmed that the difference in the MSSIM loss among the velocity components is sufficiently small ($\lesssim 0.05$).

\section{Results and discussion} \label{sec:results-discussion}

\subsection{Evaluation of the CNN inference} \label{subsec:evaluation-inference}

\begin{figure}[H]
    \centering
    \includegraphics[width=16cm]{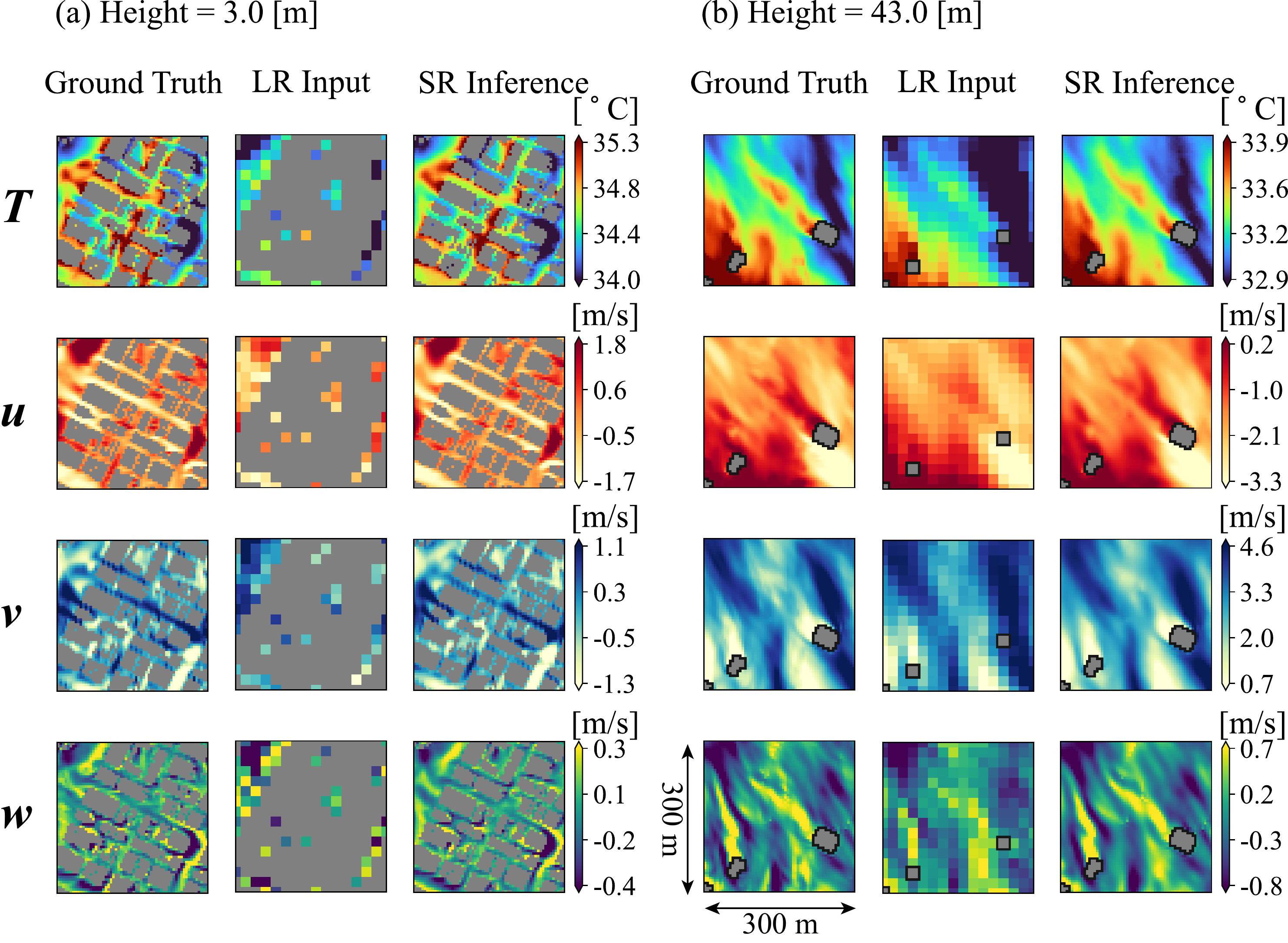}
    \caption{An example of the inference at (a) 3 m and (b) 43 m heights: air temperature ($T$) and the eastward, northward, and upward components of wind velocity ($u$, $v$, and $w$, respectively). The building grids are colored gray in the figure. In (b), the building edges are enclosed by black lines to emphasize the presence of a few buildings. The region shown here is labeled ``B'' in Fig. \ref{fig:computation-domains}.}
    \label{fig:sr-057m-z0-b}
\end{figure}

We show that the CNN successfully reconstructed HR air temperature and wind velocity fields. Figure \ref{fig:sr-057m-z0-b} shows an example of the inference at the 3 and 43 m heights. The ground elevation is not constant and its variation is incorporated in the micrometeorological simulations. For simplicity, however, this variation is ignored in the analysis because it is less than 5 m, and the height measured from the ground is used. The gray in the figure represents the missing regions within the buildings. At the 3 m height, most streets between the buildings are not represented in the LR input (Fig. \ref{fig:sr-057m-z0-b}a). Nevertheless, the CNN successfully inferred the HR temperature and velocity fields, demonstrating its ability to simultaneously perform the enhancement of spatial resolution and the reconstruction of missing data. At the 43 m height (Fig. \ref{fig:sr-057m-z0-b}b), there are fewer grid points of buildings, and the CNN shows an inference similar to the ground truth. Ordinary SR by CNNs can be interpreted as pattern matching between HR and LR images \citep{Dong2014ECCV}. Our result implies that this pattern matching allows CNNs to super-resolve three-dimensional flow fields while restoring missing data.

\begin{figure}[H]
    \centering
    \includegraphics[width=16cm]{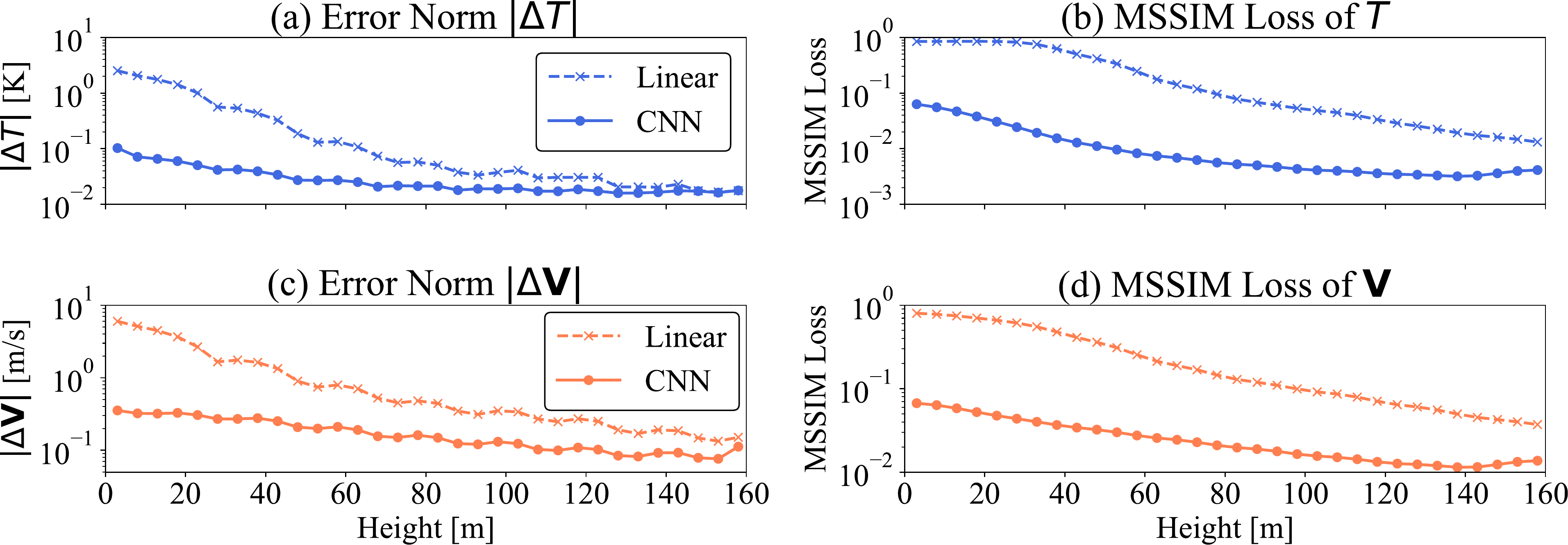}
    \caption{Dependence of the test metrics on height. The error norms and MSSIM losses are defined in (\ref{eq:mae-T}), (\ref{eq:mae-V}), and (\ref{eq:mssim}). For comparison, we employed a baseline model that uses linear interpolation and extrapolation, which is labeled ``Linear.''}
    \label{fig:error-vs-height-z0}
\end{figure}

The quantitative analysis confirmed the high accuracy of the CNN inference. Figure \ref{fig:error-vs-height-z0} shows the test metrics at all heights. As the height decreases (i.e., closer to the ground), the error norm and MSSIM loss increase, likely due to the increase in the number of building grid points. For comparison, we employed a baseline model that uses linear interpolation and extrapolation. It is obvious that this baseline model failed to infer the temperature and velocity near the ground. Indeed, the test metrics of the baseline were approximately one magnitude larger than those of the CNN (Fig. \ref{fig:error-vs-height-z0}). This result suggests that traditional methods such as linear interpolation are not capable of performing three-dimensional SR while restoring missing data. As the height increases, the inference accuracy improves for both the linear model and CNN, likely due to the decrease in the number of building grid points. Even at upper heights, the CNN was more accurate than the linear model. The test metrics of the CNN take the following maxima at the lowest (3 m) height level (Fig. \ref{fig:error-vs-height-z0}): for temperature $T$, the error norm is about 0.10 K and the MSSIM loss is about 0.06; for velocity $\bm{V}$, the error norm is about 0.35 m s$^{-1}$ and the MSSIM loss is about 0.07. Over all heights, the MSSIM loss for $T$ and $\bm{V}$ is smaller than 0.1. Such small MSSIM loss values suggest the similarity between two patterns \citep{Wang+2004IEEE}. Thus, the CNN reconstructed the three-dimensional flow patterns over the entire height range.

\begin{figure}[H]
    \centering
    \includegraphics[width=14cm]{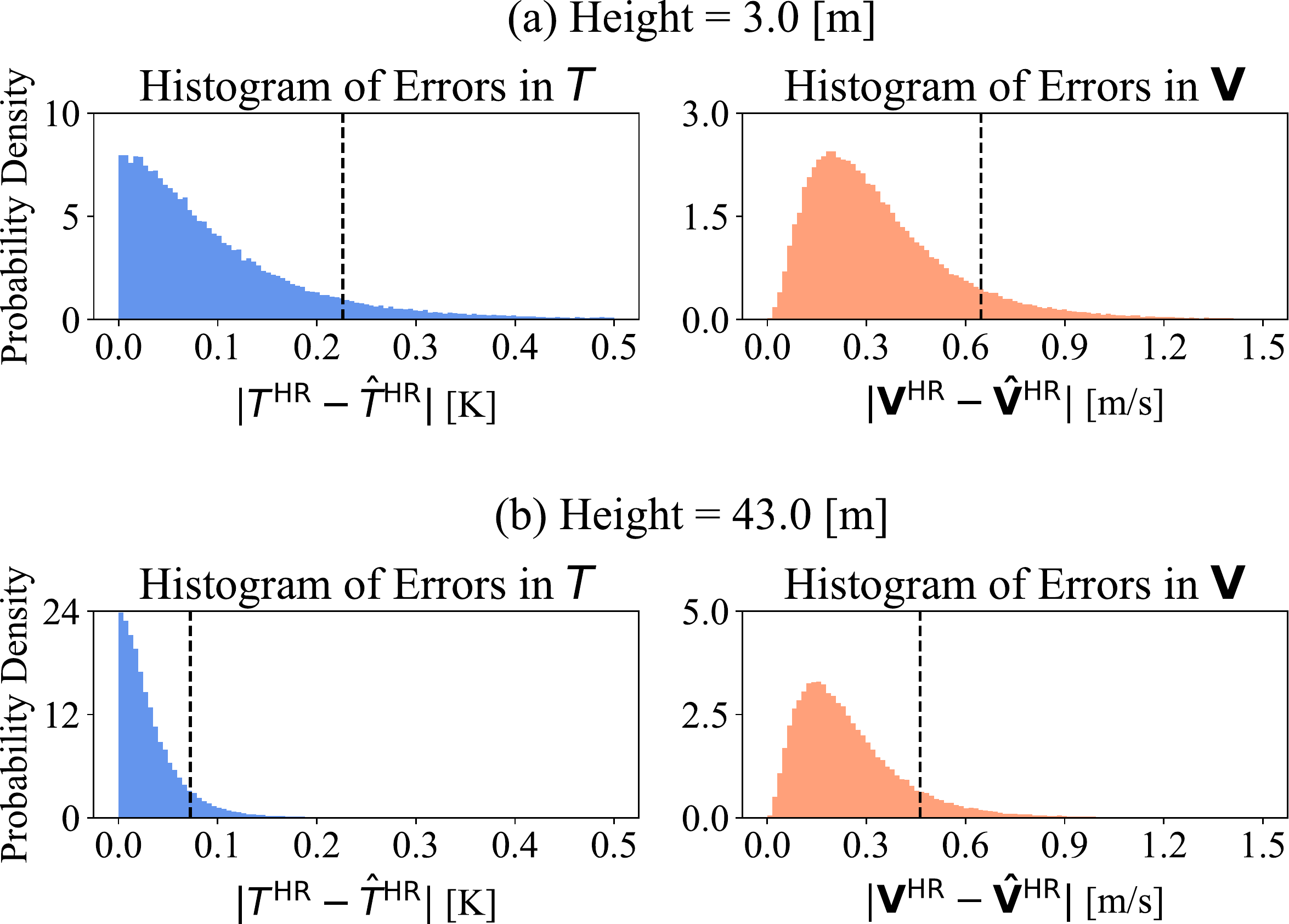}
    \caption{Histograms of the voxel-wise errors in temperature and velocity at (a) 3 m and (b) 43 m heights. The temperature error is $\left\lvert T^{\rm HR} - \hat{T}^{\rm HR} \right\rvert$ in (\ref{eq:mae-T}) and the velocity error is $\left\lvert \bm{V^{\rm HR}} - \bm{\hat{V}^{\rm HR}} \right\rvert$ in (\ref{eq:mae-V}). In each histogram, 100,000 error values were randomly sampled from the non-missing grid points in all test results. The dashed lines represent the 90 percentile values.}
    \label{fig:histogram-error-z0}
\end{figure}

Figure \ref{fig:histogram-error-z0} shows histograms of voxel-wise errors at the 3 and 43 m heights. These errors are the quantities in the summations of (\ref{eq:mae-T}) and (\ref{eq:mae-V}). Specifically, the temperature error is given by the absolute value of the temperature difference, i.e., $\left\lvert T^{\rm HR} - \hat{T}^{\rm HR} \right\rvert$ in (\ref{eq:mae-T}), which has peaks around zeros in Fig. \ref{fig:histogram-error-z0}. The velocity error is equal to the Euclidean norm of the difference in velocity vectors, i.e., $\left\lvert \bm{V^{\rm HR}} - \bm{\hat{V}^{\rm HR}} \right\rvert$ in (\ref{eq:mae-V}), which has peaks slightly away from zeros in Fig. \ref{fig:histogram-error-z0} because the error consists of the sum of squared quantities. The 3 m height is closest to the ground, resulting in wider error distributions compared to 43 m. At the 3 m height, the 90-percentile errors are 0.23 K for $T$ and 0.64 m s$^{-1}$ for $\bm{V}$, while at the 43 m height, 0.07 K for $T$ and 0.46 m s$^{-1}$ for $\bm{V}$. The scaling parameters in (\ref{eq:preprocess}) are 8.4 K for $T$ and 14.3 m s$^{-1}$ for $\bm{V}$. The latter is the average of the parameters for $u$, $v$, and $w$. Even at the 3 m height, the 90-percentile errors are 2.7\% and 4.5\% of these scales for temperature and velocity, respectively. Thus, the voxel-wise error in the CNN inference is sufficiently small.

\begin{figure}[H]
    \centering
    \includegraphics[width=14cm]{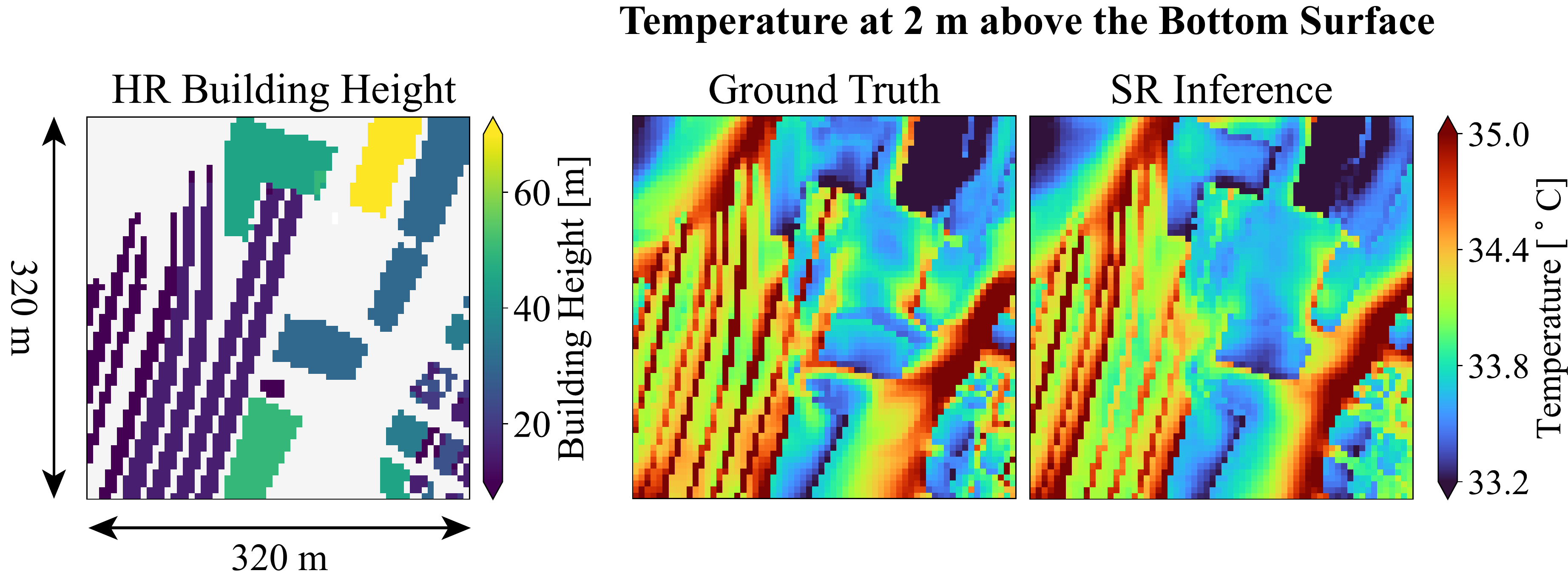}
    \caption{An example of the two-dimensional temperature field at the 2 m height above the bottom surface (e.g., the ground or roofs of buildings). The region shown here is located at Tokyo Station, where multiple railway platforms are present. This region is labeled ``A'' in Fig. \ref{fig:computation-domains}. Our previous study shows an inference at the same location (Fig. 6 in \citep{Yasuda2022BAE}).}
    \label{fig:2m-height-temperature}
\end{figure}

The accuracy of temperature inference near the ground is comparable to that of our previous study \citep{Yasuda2022BAE}. 
The present error norm was 0.10 K at the 3 m height (Fig. \ref{fig:error-vs-height-z0}), while the root-mean-square error (RMSE) was 0.15 K for the 2 m height in our previous study \citep{Yasuda2022BAE}. In that study, two-dimensional temperatures were generated by sampling grid points at 2 m above the bottom surface (e.g., the ground or roofs of buildings). In such a manner of creating two-dimensional data, no missing value appeared. For direct comparison, two-dimensional temperature fields were obtained in the same manner from both the ground truth and CNN inference. Figure \ref{fig:2m-height-temperature} shows an example of the resulting temperature. The region shown in the figure is located at Tokyo Station and is characterized by multiple railway platforms, as seen in the building height distribution. The temperature is lower at points of taller buildings because the temperatures above these buildings were sampled, leading to the discontinuities in temperature (Fig. \ref{fig:2m-height-temperature}). The CNN reproduced these discontinuities, which indicates the accuracy of the base, three-dimensional inference. Furthermore, the mean absolute error of the two-dimensional temperature is 0.13 K, whereas the corresponding error was 0.11 K in our previous study \citep{Yasuda2022BAE}. The proposed CNN exhibits comparable accuracy, even though it is not specifically designed to infer the two-dimensional temperature.

The computation time for the CNN inference is negligibly small, implying that the CNN can be employed as an NN in SR simulation systems \citep{Onishi+2019SOLA, Wu+2021GRL, Wang+2021GMD}. For instance, it took approximately $0.27$ seconds to generate an inference on the full size of $320 \times 320 \times 32$ voxels. The wall time for SR simulations can be estimated as follows. An LR micrometeorological simulation for 1-hour predictions would be conducted, which generates 60 sets of 1-minute average data. The elapsed time of this simulation was estimated at 312 seconds \citep{Yasuda2022BAE}. The CNN would then need 16.2 ($=0.27 \times 60$) seconds to super-resolve all LR results. Thus, the total wall time for the SR simulation would be about 328 seconds. This estimated time is much shorter than the elapsed time for the HR micrometeorological simulation, which is approximately 8 hours, highlighting the computational efficiency of SR simulation.

\subsection{Inference only utilizing upper flows} \label{subsec:evaluation-only-upper}

We demonstrate that near-surface flows can be inferred from upper flows. As previously discussed, the simultaneous enhancement of spatial resolution and restoration of missing regions can be interpreted as pattern matching between LR and HR flows. In the vicinity of the ground, the LR input contains many missing grid points due to the presence of buildings (Fig. \ref{fig:sr-057m-z0-b}a). In contrast, at upper levels, such missing points become fewer (Fig. \ref{fig:sr-057m-z0-b}b). This result suggests that flows at upper levels are more informative for pattern matching. In addition, the spatial scale of upper flows is larger due to the absence of buildings, compared to near-surface flows. This implies that the small-scale flows near the ground can be inferred from the large-scale upper flows. To examine this hypothesis, we made all grid points missing below a certain height in the LR input and conducted the training and evaluation of the CNN.

\begin{figure}[H]
    \centering
    \includegraphics[width=9cm]{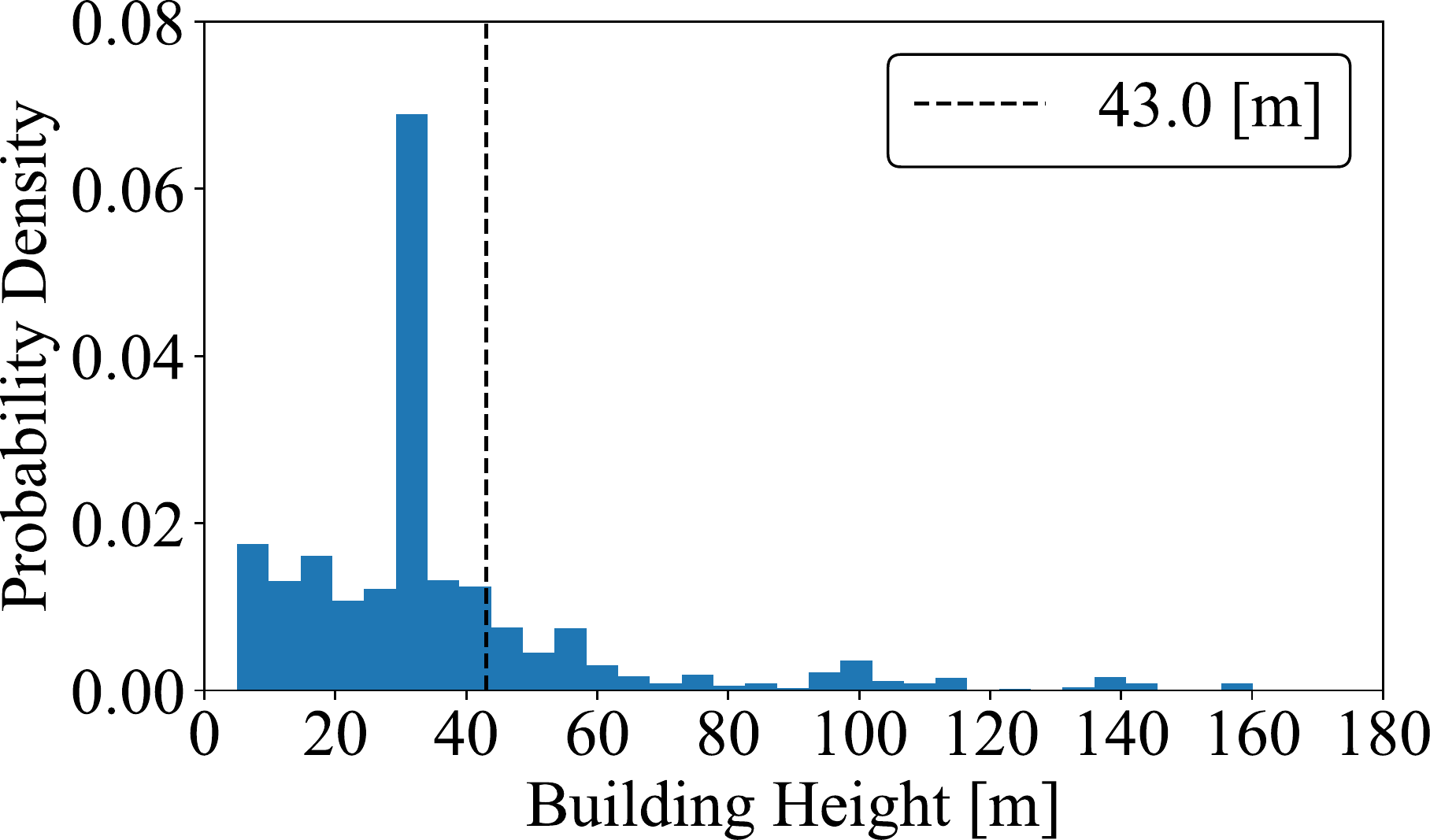}
    \caption{Histogram of the building height around Tokyo Station. The corresponding building-height spatial distribution is shown in Fig. \ref{fig:computation-domains}.}
    \label{fig:histogram-building-height}
\end{figure}

We first determined an appropriate height threshold for the replacement with missing values. Figure \ref{fig:histogram-building-height} shows a histogram of the building height. The peak of the histogram is around 30 m, which is due to the height cap of 31 m in the old Building Standards Law \citep{Song2020}. The figure indicates that most buildings are lower than 43 m in height. Thus, we replaced all values below 43 m in the LR input with the missing values, i.e., zeros.

\begin{figure}[H]
    \centering
    \includegraphics[width=14cm]{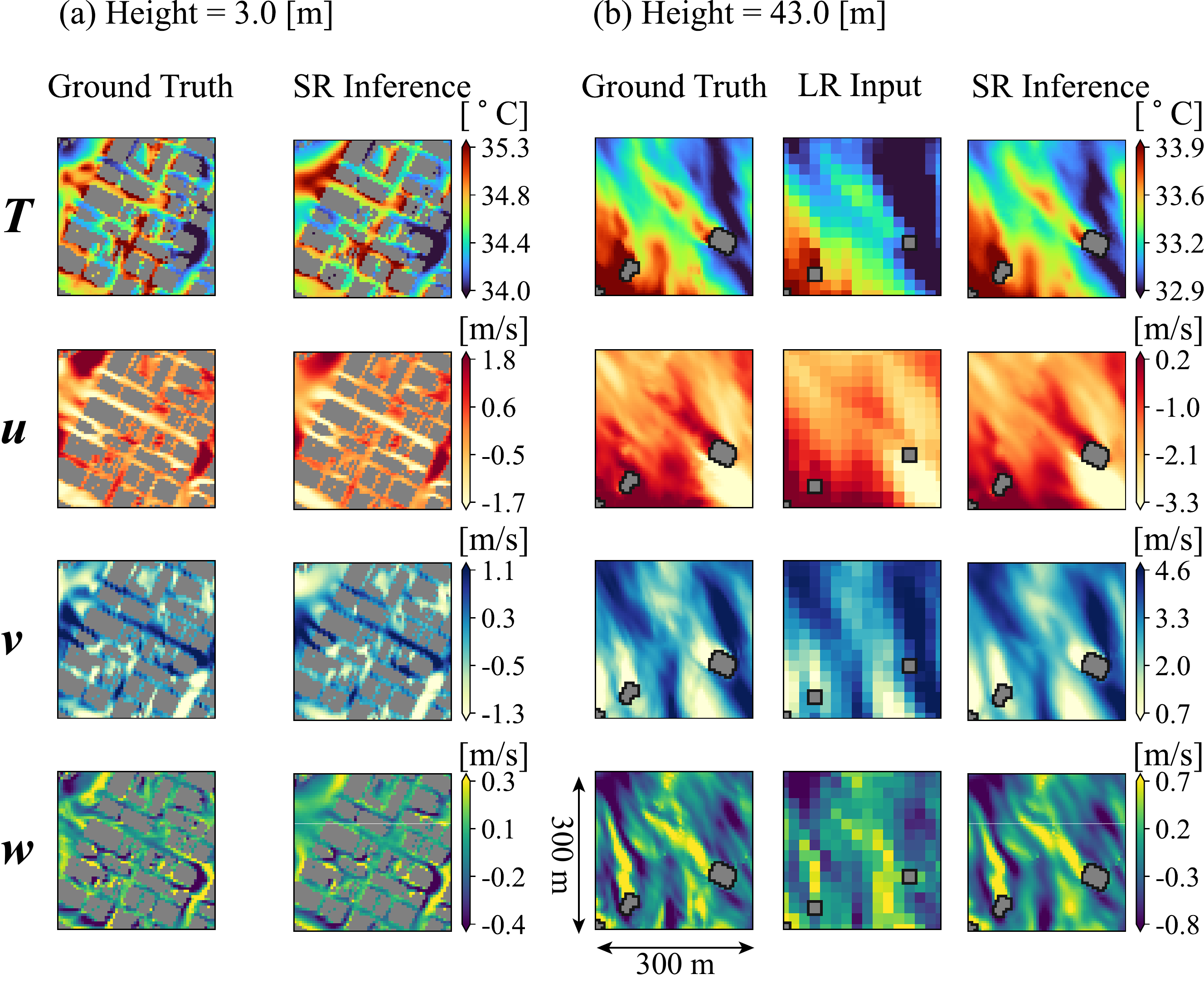}
    \caption{An example of the CNN inference at (a) 3 m and (b) 43 m heights: air temperature ($T$) and the eastward, northward, and upward components of wind velocity ($u$, $v$, and $w$, respectively). All input values below the 43 m height were made missing and the CNN was then trained and evaluated. The LR input is not shown in (a), as all values are missing. The building grids are colored gray in the figure. In (b), the building edges are enclosed by black lines to emphasize the presence of a few buildings. The region shown here is the same as in Fig. \ref{fig:sr-057m-z0-b} and is labeled ``B'' in Fig. \ref{fig:computation-domains}.}
    \label{fig:sr-057m-z2-b}
\end{figure}

The CNN can qualitatively infer near-surface flows from upper flows, as demonstrated in Fig. \ref{fig:sr-057m-z2-b}. The LR input is not shown at the 3 m height (Fig. \ref{fig:sr-057m-z2-b}a) since all values are missing. Despite these missing data, the CNN reproduced the flow structures that were qualitatively similar to those of the ground truth. In particular, the CNN accurately identified the locations of small or large values of temperature and velocity, although the detailed patterns differed from those of the ground truth. At the 43 m height (Fig. \ref{fig:sr-057m-z2-b}b), no replacement with the missing values was performed in the LR input, which is likely to lead to a more accurate inference, compared to Fig. \ref{fig:sr-057m-z2-b}a.

\begin{figure}[H]
    \centering
    \includegraphics[width=16cm]{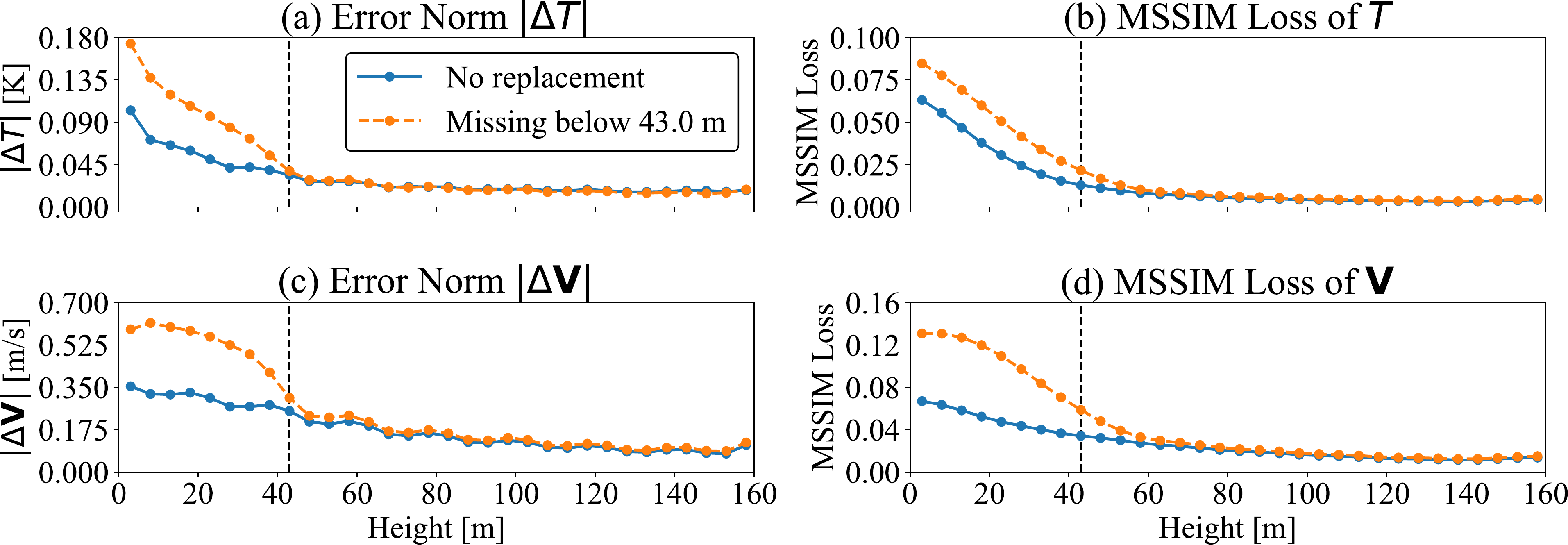}
    \caption{Dependence of the test metrics on height. The error norms and MSSIM losses are defined in (\ref{eq:mae-T}), (\ref{eq:mae-V}), and (\ref{eq:mssim}). The CNN was trained and evaluated using the LR input where all values below the 43 m height were made missing. This height threshold is denoted by the dashed vertical lines. The obtained result is labeled ``Missing below 43.0 m.'' The result of the original CNN is denoted by ``No replacement,'' whose error values are the same as in Fig. \ref{fig:error-vs-height-z0}.}
    \label{fig:error-vs-height-z0-vs-z2}
\end{figure}

Figure \ref{fig:error-vs-height-z0-vs-z2} compares the accuracy of inference with and without the missing-value replacement. The case without the replacement has been examined in Fig. \ref{fig:error-vs-height-z0}. The vertical dashed lines in Fig. \ref{fig:error-vs-height-z0-vs-z2} represent the 43 m height. Above this threshold, the test metrics are comparable for temperature and velocity with and without the missing-value replacement. However, at lower heights, the error norm and MSSIM loss are larger in the case of the missing-value replacement. In particular, at the 3.0 m height (i.e., the bottom-most level) this replacement increased the error norm of temperature from 0.10 to 0.17 K and the MSSIM loss from 0.06 to 0.09, whereas for velocity, the error norm increased from 0.35 to 0.59 m s$^{-1}$ and the MSSIM loss from 0.07 to 0.13. Although the error norm and MSSIM loss are approximately 1.5 to 2.0 times larger, the MSSIM loss still remains close to 0.1, leading to the MSSIM being around 0.9 according to (\ref{eq:mssim}). When the MSSIM is greater than about 0.9, two images are perceptually similar \citep{Wang+2004IEEE}. Therefore, we conclude that near-surface flows can be qualitatively inferred by the CNN from flows at upper heights.

\section{Conclusions} \label{sec:conclusions}

This study has proposed a convolutional neural network (CNN) that super-resolves three-dimensional air temperature and wind velocity fields obtained from building-resolving micrometeorological simulations. The CNN architecture is based on a U-Net for image inpainting \citep{Schweri+2021FC}. The super-resolution (SR) process in the CNN requires not only increasing the spatial resolution but also restoring the missing regions caused by buildings because the building shape is different between low and high resolution (LR and HR). The proposed CNN incorporates an image inpainting technique, called gated convolution \citep{Yu+2019ICCV}. Additionally, by clipping the input to the range of $[0, 1]$, zeros are regarded as missing values. This preprocessing eliminates the need for LR building information and allows the CNN to perform inference using only the HR building data.

The CNN performance has been evaluated via supervised learning using the results of micrometeorological simulations for an actual urban area around Tokyo Station in Japan. The LR input data were created by applying the average pooling to the HR simulation results. The training loss included a physics loss term, namely the error in the divergence of velocity vectors. The CNN reconstructed the velocity and temperature fields with high accuracy. The error was larger near the ground than at upper heights, which is likely due to the missing data in the input caused by the coarseness of the LR buildings. The mean errors at the lowest height were approximately 0.10 K for temperature and 0.35 m s$^{-1}$ for velocity. Furthermore, we have demonstrated that small-scale near-surface flows can be qualitatively inferred from the large-scale flows above most buildings. This result implies that flows around buildings can be estimated from the LR simulation results in which buildings are not accurately represented.

There are at least two potential directions for future research. The first direction is to investigate the generalizability of the proposed CNN to other cities. Our previous study \citep{Yasuda2022BAE} demonstrated that the SR model for the two-dimensional temperature can be applied to a city different from the city used in training. It would be interesting to examine whether similar results are obtained with the proposed CNN for the three-dimensional SR. The second direction is to explore the generation of LR and HR data from separate simulations, since in the concepts of SR simulation \citep{Onishi+2019SOLA, Wu+2021GRL, Wang+2021GMD}, LR and HR data are obtained from LR and HR atmospheric simulations, respectively, and a neural network is trained using these data. The effectiveness of SR simulation has been shown on meteorological or climate scale \citep{Wu+2021GRL, Wang+2021GMD}. In urban micrometeorology, however, it is necessary to incorporate changes in building shapes during the SR process. Based on the proof of concept in this paper, it may be possible to develop an SR simulation system for real-time micrometeorological predictions in urban cities.

\section*{Declaration of competing interest}

The authors declare that they have no known competing financial interests or personal relationships that could have appeared to influence the work reported in this paper.

\section*{Data availability}

The source code for deep learning is preserved at the Zenodo repository (\url{https://doi.org/10.5281/zenodo.7777210}) and developed openly at the GitHub repository (\url{https://github.com/YukiYasuda2718/3d-sr-micrometeorology}). The data that support the findings of this study are available from the corresponding author upon reasonable request.

\section*{Acknowledgements}

This work was supported by the JSPS KAKENHI (grant number 20H05751). The micrometeorological simulation and deep learning were performed on the Earth Simulator system (project ID: 0-22021) at the Japan Agency for Marine-Earth Science and Technology (JAMSTEC).



\bibliographystyle{elsarticle-num} 
\bibliography{references}





\end{document}